# Effect of Molybdenum 4*d* Hole Substitution in BaFe$_2$As$_2$

Athena S. Sefat,[*] Karol Marty,[*] Andrew D. Christianson,[*] Bayrammurad Saparov,[*] Michael A. McGuire,[*] Mark D. Lumsden,[*] Wei Tien[∇], Brian C. Sales[*]

[*]*Oak Ridge National Laboratory, Oak Ridge, Tennessee 37831 USA*
[∇]*Ames National Laboratory and Department of Physics and Astronomy, Iowa State University, Ames, Iowa 50011, USA*

**Abstract**

We investigate the thermodynamic and transport properties of molybdenum-doped BaFe$_2$As$_2$ (122) crystals, the first report of hole doping using a 4*d* element. The chemical substitution of Mo in place of Fe is possible up to ~ 7%. For Ba(Fe$_{1-x}$Mo$_x$)$_2$As$_2$, the suppression rate of the magnetic transition temperature with x is the same as in 3*d* Cr-doped 122 and is independent of the unit cell changes. This illustrates that temperature-composition phase diagram for hole-doped 122 can be simply parameterized by x, similar to the electron-doped 122 systems found in literature. Compared to 122 with a coupled antiferromagnetic order (T$_N$) and orthorhombic structural transition (T$_0$) at ≈ 132 K, 1.3% Mo-doped 122 (x = 0.013) gives T$_N$ = T$_0$ = 125(1) K according to neutron diffraction results and features in specific heat, magnetic susceptibility and electrical resistivity. The cell volume expands by ~ 1% with maximum Mo-doping and T$_N$ is reduced to ≈ 90 K. There is a new T$^*$ feature that is identified for lightly Cr- or Mo-doped (< 3%) 122 crystals, which is x dependent. This low-temperature transition may be a trace of superconductivity or it may have another electronic or magnetic origin.

**Introduction**

High-temperature superconductivity continues to attract attention since its causes in both Fe-based superconductors (FeSCs) and cuprates remain unsolved. In each, the superconducting state is produced by suppression of the magnetic phase transition of an antiferromagnetic 'parent' material. The FeSC parents are itinerant and only weakly correlated, in contrast to the Mott-insulating cuprate parents. Also distinct from the cuprates, certain applications of pressure [1] or chemical substitutions in the Fe-plane [2, 3] of FeSC parents instigate superconductivity.

BaFe$_2$As$_2$ (122) is one of FeSC parents, with a room temperature ThCr$_2$Si$_2$-type tetragonal crystal structure. It has a spin-density-wave (SDW) order below ≈ 132 K, where Fe spins are aligned antiferromagnetically along *a*- and *c*-axes, and ferromagnetically along *b*-axis [4, 5]. This magnetic transition temperature (T$_N$) is coupled with an orthorhombic phase (T$_o$) transition [4, 5]. In 122, superconductivity can be tuned by electron doping the Fe-site using other 3*d* (Co, Ni) or 4*d* (Rh, Pd) transition metals (*TM*) in Ba(Fe$_{1-x}$*TM*$_x$)$_2$As$_2$ systems [2, 6]. For Co- and Rh-doping or Ni- and Pd-doping, the rate of T$_N$ suppression, the maximum superconducting transition temperature (T$_C$), and the range of superconducting dome are found to be identical [7]. Thus, 3d and 4d dopants belonging to a group in the periodic table produce overlapping temperature-composition (T-x) phase diagrams. Although this emphasizes that changes in unit cell dimensions may not be significant, electron count is important as 2.5% Ni- or 5% Co-doping of 122 produces optimal $T_C$ [7]. Compared to such electron doped systems, hole doping on the Fe-site using 3d (Cr, Mn) also give similar $T_N$ suppression rates with x, but new magnetic phases are stabilized instead of superconductivity [8, 9, 10]. In fact for all $Ba(Fe_{1-x}Cr_x)_2As_2$ concentrations of x ≤ 0.47, long-range antiferromagnetic order is observed [8, 9]: for x < 0.3, it is the SDW C-type, same as 122; for x ≥ 0.3, it is G-type where Fe moments are aligned antiferromagnetically along all crystal axes. These results are argued to be a consequence of strong spin dependent hybridization between Cr (or Mn) d- and As p-states, with substitutions in $BaFe_2As_2$ resulting in strong scattering and carrier localization [8, 11]. In contrast, modest covalency between Fe (or Co, or Ni) with As states were found [2, 11].

Here we study the properties of $Ba(Fe_{1-x}Mo_x)_2As_2$, a first report of hole doping of 122 using a 4d element. Compared to $Fe^{2+}$ with $3d^6$, $Mo^{2+}$ has two less valence electrons ($4d^4$). The possibility of finding superconductivity in $Ba(Fe_{1-x}Mo_x)_2As_2$ or having a similar T-x phase diagram to the already reported Cr-doped 122 [8, 9] are the aims of study here. We present results from x-ray and neutron diffraction, and thermodynamic and transport properties.

**Results and Discussions**

The single crystals of $Ba(Fe_{1-x}Mo_x)_2As_2$ (Mo-122) were grown out of a mixture of Ba, FeAs flux, and Mo or MoAs. High purity elements (>99.9%, from Alfa Aesar) were used. FeAs binary was synthesized by heating elemental iron and arsenic in a sealed silica crucible to 350 °C (50 °C/hr, hold 2 hrs), to 600 °C (20 °C/hr, hold 10 hrs), then to 1060 °C (40 °C/hr, hold 8 hrs), ended by furnace cooling. We bound Mo with As by heating the 1:1 ratio of the elements, in a sealed silica crucible, to 350 °C (50 °C/hr, hold 5 hrs), to 600 °C (30 °C/hr, hold 20 hrs), to 650 °C (20 °C/hr, hold 40 hrs), then furnace cooling. Various loading ratios of Ba:Mo(or MoAs):FeAs, with Ba = 1, were used to produce the variety of chemical substitutions, listed in Table 1. Each of these mixtures was heated for ~ 20 hrs at 1200 °C, and then cooled at a rate of 1 to 2 °C/hr, followed by decanting of the flux at 1100 °C or 1120 °C. Mo-122 crystals had sheet morphologies and dimensions of ~ 5 × 4× 0.1 $mm^3$ or smaller in a, b, and c crystallographic directions, respectively. Similar to $BaFe_2As_2$ [2], the crystals formed with the [001] direction perpendicular to the plane of the plate.

Attempts of crystal growths for higher Mo content than those presented in Table 1 were unsuccessful, producing a mixture of phases with $Mo_{5-y}Fe_yAs_4$ with 0 < y < 1, according to EDS results. The limitation for Mo substitution on the Fe-site (less than 7%) may be due to the formation of such Mo-rich phases and partial solubility of Mo in FeAs solution. In an attempt to solve the latter, we used MoAs nominal compound in FeAs solution, but this similarly failed to give higher Mo-doping. For loading ratio of FeAs:Mo=5:1.5 and those with higher Mo content, secondary $Mo_{5-y}Fe_yAs_4$ phases were obtained, in addition to Mo-122 crystals. For these, the yield and size of the latter were smaller. In contrast to Mo-122, Cr substitutions in Fe-site was possible up to ~ 45% in $Ba(Fe_{1-x}Cr_x)_2As_2$ [8, 9]. We have made seven new Cr-doped 122 crystals, and up to 61% doping here, following the synthesis procedures in the original report [5].

Chemical compositions of the crystals were found using a Hitachi S3400 Scanning Electron Microscope operating at 20 kV. The beam current was set to provide approximately 1500 counts/second using a 10 mm sq EDAX detector set for a processing time of 54 microseconds. Data were reduced using Energy-dispersive X-ray spectroscopy (EDS) Standardless Analysis Program. The EDS analysis indicated that a lot less molybdenum was chemically substituted in 122 crystals than put in solution; these results are summarized in Table 1. Three spots were averaged on the surface of each single crystal. The samples are denoted by be these measured EDS x values in Ba(Fe$_{1-x}$Mo$_x$)$_2$As$_2$ throughout this paper.

The phase purity of the Ba(Fe$_{1-x}$Mo$_x$)$_2$As$_2$ crystals was checked by collecting data on an X'Pert PRO MPD X-ray powder diffractometer in 5-90° 2$\theta$ range. At room temperature, the structures were identified as the tetragonal ThCr$_2$Si$_2$ structure-type (*I4/mmm*, $Z = 2$). The air and moisture stability of the materials were confirmed by rechecking the diffraction scan of a sample left overnight. Lattice parameters were determined from full-pattern LeBail refinements using X'Pert HighScore. For BaFe$_2$As$_2$, $a = 3.9619(2)$ Å and $c = 13.0151(5)$ Å. With Mo-doping, the cell volume increases, for example for x = 0.013, it increases by 0.05%, where $a = 3.9607(3)$ Å and $c = 13.0298(8)$ Å. For x = 0.049, the cell volume increases by 0.50%, where $a = 3.9598(2)$ Å and $c = 13.0948(7)$ Å. Figure 1a plots *a*- and *c*-lattice parameters as a function of Mo concentration (x). With Mo-doping, although *a*-lattice parameter reduces weakly, *c*-lattice parameter expands strongly (also see Table 1). Because of the expected larger size of Mo compared with Cr, Cr-122 gives less notable change in *c*-lattice parameter, although *a*-lattice parameter increases softly in the same range of x (Fig. 1b, inset). In Ba(Fe$_{1-x}$Cr$_x$)$_2$As$_2$, 5% Cr-doping expands the cell volume by 0.34% mainly due to the increase in the *c*-lattice. In place of Fe in 122, ~ 7% or ~ 60% chemical substitution of Mo or Cr, respectively, produces an overall unit cell volume expansion of ~ 1% or ~ 3%.

Table 1: For flux-solution grown Ba(Fe$_{1-x}$Mo$_x$)$_2$As$_2$ crystals, the loading ratio of transition metal elements are listed. The resulting molybdenum concentration (x) and the refined *c*-lattice parameters are listed.

| Loading amount | Mo (x) | *c*-lattice parameter (Å) |
|---|---|---|
| FeAs:Mo=5:0.1 | 0.006 | 13.0195(7) |
| FeAs:MoAs=4.9:0.1 | 0.009 | 13.0208(7) |
| FeAs:Mo=5:0.25 | 0.013 | 13.0298(8) |
| FeAs:Mo=5:0.5 | 0.026 | 13.0484(6) |
| FeAs:MoAs=4.5:0.5 | 0.028 | 13.0517(8) |
| FeAs:Mo=5:1 | 0.045 | 13.0740(8) |
| FeAs:Mo=5:1.5 | 0.049 | 13.0948(7) |
| FeAs:Mo=5:1.6 | 0.054 | 13.112(2) |
| FeAs:Mo=5:2.2 | 0.066 | 13.120(1) |

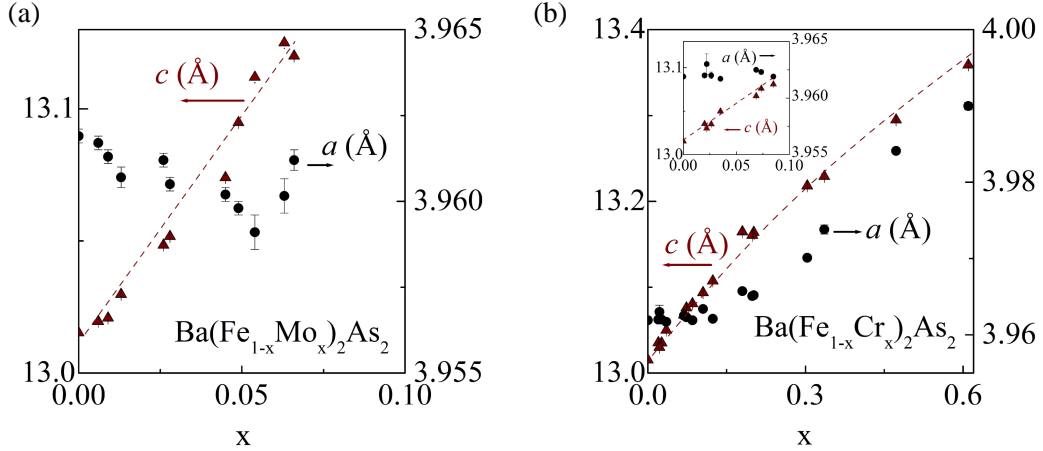

Fig. 1: Room-temperature tetragonal lattice parameters for the range of possible chemical substitutions in Ba(Fe$_{1-x}$Mo$_x$)$_2$As$_2$ (a) and Ba(Fe$_{1-x}$Cr$_x$)$_2$As$_2$ (b).

Magnetization results for Ba(Fe$_{1-x}$Mo$_x$)$_2$As$_2$ were collected using a Quantum Design Magnetic property Measurement System with applied field along *ab*-crystallographic direction. For a typical temperature-sweep experiment, the sample was cooled to 1.8 K in zero field (zfc) and data were collected by warming from 1.8 K to ~ 380 K, in 1 Tesla. The data are presented in log plot of $\chi/\chi_{380K}$ for selected x, in Figure 2a. For BaFe$_2$As$_2$, $\chi_{ab}$(300K) ≈ 0.001 cm$^3$/mol. Similar to Cr-122 [8], Mo doping increases the magnitude of the magnetic susceptibility at room temperature; opposite effect is seen for Co-122 [12]. For x = 0.066, $\chi_{ab}$(300 K) reaches ≈ 0.005 cm$^3$/mol. For x = 0, $\chi$ decreases linearly with decreasing temperature and drops abruptly below $T_N$ = $T_0$ ≈ 132 K, reproducing the result seen in literature [4, 5]. For x ≤ 0.028, the behavior of $\chi$(T) is the same, although the transition temperatures are reduced with x. For x ≥ 0.045, $\chi$ values increase with decreasing temperature and give further reductions in the transition temperatures. In order to infer $T_N$, the data were converted to Fisher's d($\chi$T)/dT [13]. For x = 0.006, 0.009, 0.013, 0.026, 0.028, 0.045, 0.049, and 0.066, $T_N$ values are ≈ 128 K, 127 K, 124 K, 120 K, 118 K, 109 K, 98 K, and 89 K, respectively. The rate of decrease of $T_N$ with Mo-122 is similar to that seen for Cr-122 [8].

Field-dependent magnetization, M(H), at 2 K was found approximately linear for x ≤ 0.049 and up to 6.8 Tesla (Fig. 2a, inset). There are small non-linear regions below ~ 0.2 Tesla for x = 0 and 0.013, and below ~ 0.4 Tesla for x = 0.028 and 0.049. For higher Mo-doped crystals such as x = 0.066, there is a small hysteresis with no saturation up to 6.8 Tesla and remanent magnetization estimated at $M_r$ = 0.015 $\mu_B$/f.u.

Magnetization data were also collected at a smaller field of 0.1 Tesla, in zfc and field-cooled (fc) conditions, for BaFe$_2$As$_2$ and a lightly Mo-doped x = 0.013 (Fig. 2b). Although zfc/fc data overlap for x = 0 (inset), there is a weak splitting between zfc and fc data for x = 0.013 that is field dependent. This feature was reproduced on several crystals of the same batch and it hints of a small ferromagnetic component. A much larger ferromagnetic moment is observed in strained SrFe$_2$As$_2$ in literature, with hysteretic behavior of M(H) that saturates near M(H) [14]. In contrast to this, M(H) is approximately linear here for Mo-122 with x ≤ 0.049 and seems x-dependent.

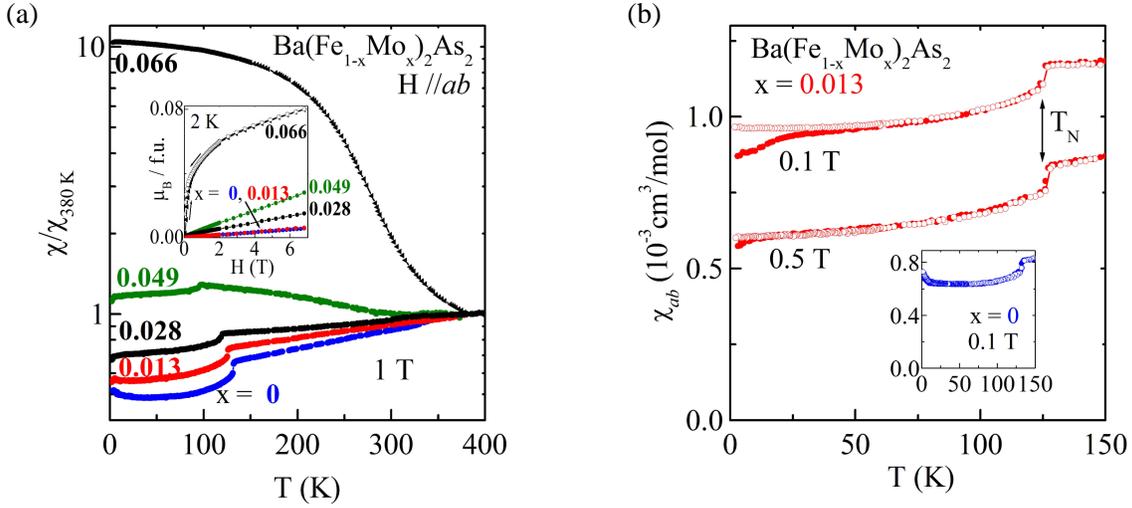

Fig. 2: Magnetization data for selected x in Ba(Fe$_{1-x}$Mo$_x$)$_2$As$_2$, measured along the *ab*-crystal direction. (a) The zero-field cooled (zfc) temperature dependence of the scaled molar susceptibility in 1 Tesla; inset is the field dependence of magnetization at 2 K. (b) The zfc and field cooled (fc) temperature dependence of molar susceptibility of x = 0.013, measured in 0.1 Tesla and 0.5 Tesla; inset is the overlapping zfc and fc data for BaFe$_2$As$_2$ measured in 0.1 Tesla.

Transport measurements for Ba(Fe$_{1-x}$Mo$_x$)$_2$As$_2$ were performed with a Quantum Design Physical Property Measurement System (PPMS). Electrical platinum leads were attached onto the sample using Dupont 4929 silver paste and resistance measured in the *ab* plane in the range of 1.8 K to 350 K. The ρ values at 300 K ranged from 0.5 to 1.8 mΩ cm, although their absolute values may have suffered from the geometry factor estimations.

Figure 3a presents scaled data in the form of ρ/ρ$_{350K}$. Electrical resistivity for x = 0 diminishes with decreasing temperature from 350 K, falling rapidly below 134 K associated with T$_N$ and T$_o$ [4, 5]. For lightly Mo-doped compositions of 0.006 ≤ x ≤ 0.028, ρ(T) first reduces gradually, then enhances weakly to a sharp feature (≈ T$_N$ = T$_0$), followed by a steady decrease below, and finally a rapid drop below T$^*$ (see Fig. 3 insets). The resistive transition temperatures can be estimated from peaks in dρ/dT [15]. The T$_N$ and T$^*$ approximate values, respectively, are 128 K and 16.3 K for x = 0.006, 127 K and 16.7 K for x = 0.009, 124 K and 12.5 K for x = 0.013, 120 K and 7.4 K for x = 0.026, and 118 K and 4.9 K for x = 0.028. The resistivity for x ≥ 0.045 first decreases gently from room temperature, followed by sharp upturns below ~ 100 K probably associated with changes in the scattering due to the decrease in carriers. For these higher Mo-doped crystals, the T$_N$ values estimated from dρ/dT are ≈ 106 K for x = 0.045, 106 K for x = 0.049, 98 K for x = 0.054, 91 K for x = 0.063, and 89 K for x = 0.066.

No thermal hysteresis was observed in the ρ$_{ab}$(T) data, suggesting that neither of the T$_N$ or T$^*$ transitions are strongly first ordered. Figure 3b shows the ρ(T) data for x = 0.013 and x = 0.049 in an applied magnetic field. Although T$_N$ does not shift in field for x = 0.013 and x = 0.049, T$^*$ shifts in x = 0.013 from T$^*_{onset}$ (0T) = 14.7 K to T$^*_{onset}$ (10T) = 10.9 K. The T$^*$ transition may be a result of filamentary superconductivity. Traces of superconductivity are seen in parents of SrFe$_2$As$_2$ and BaFe$_2$As$_2$ in literature and are argued to be a result of residual strain [14, 16]. Although crystal strain is a possibility here, it does not explain the composition (x) dependence of T$^*$.

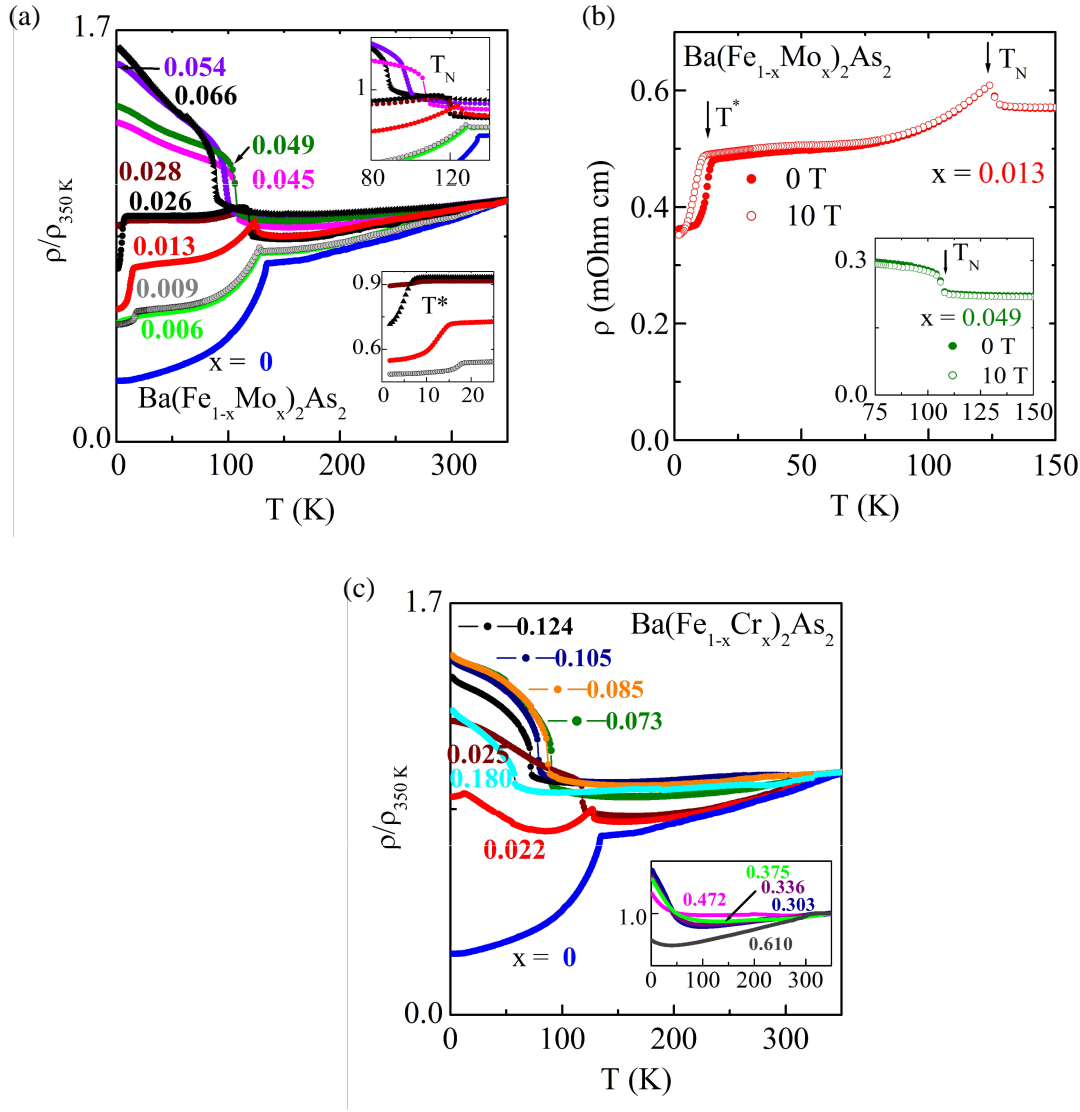

Fig. 3: (a) For Ba(Fe$_{1-x}$Mo$_x$)$_2$As$_2$, temperature dependence of electrical resistivity measured in the *ab*-plane for the range of 0 ≤ x ≤ 0.066. Top inset illustrates the features related to the structural and magnetic SDW (T$_N$) transition. Bottom inset illustrates the anomaly in resistivity at lower temperatures (T$^*$) that appears for the lightly Mo-doped compositions. (b) For Ba(Fe$_{1-x}$Mo$_x$)$_2$As$_2$ with x = 0.013 and x = 0.045, the effects of 10 Tesla field on antiferromagnetic (T$_N$) and low-temperature (T$^*$) transitions. (c) For Ba(Fe$_{1-x}$Cr$_x$)$_2$As$_2$, temperature dependence of electrical resistivity measured in the *ab*-plane for 0 ≤ x ≤ 0.610.

In order to evaluate 4$d$ Mo-122 in relation to 3$d$ Cr-122, we synthesized and measured $\rho(T)$ for a wide range of compositions in Ba(Fe$_{1-x}$Cr$_x$)$_2$As$_2$; see Figure 3c. Compared to the original report [8], here we show data for seven additional compositions and find two new features. One is that for semiconducting behavior is observed for $0.303 \leq x \leq 0.472$ (Fig. 3b, inset). For this composition range, neutron diffraction gives G-type antiferromagnetic order [9]. Considering this finding, it may be interesting to see if the magnetic ground state changes for $x \geq 0.610$ that show metallic behavior. Two is that for the lightly Cr-doped composition of x= 0.022, in addition to the T$_N$ feature at 125 K, there is a second feature at T$^*$ ≈ 10.9 K in d$\rho$/dT. Thus, the low temperature T$^*$ anomaly is commonly observed in both the hole-doped 3$d$ and 4$d$, and although it does not reach zero down to 1.8 K, it may be a trace of superconductivity. Otherwise, it may have another electronic or magnetic origin.

Although the Hall coefficient ($R_H$) is negative for temperature range of ~ 150 K to 300 K for x = 0 [17] and for x < 0.07 in Ba(Fe$_{1-x}$Cr$_x$)$_2$As$_2$ [8], its tendency to move towards more positive values with x is an indication of Cr acting as a hole dopant in 122 [8]. Even though we have not measured $R_H$ in Ba(Fe$_{1-x}$Mo$_x$)$_2$As$_2$, we expect for them to act similarly.

Specific heat data were collected on Ba(Fe$_{1-x}$Mo$_x$)$_2$As$_2$ single crystals, also using a PPMS, see Figure 4. For BaFe$_2$As$_2$, a sharp transition is observed at 131.6 K, associated with the T$_0$ and T$_N$ [4]. With Mo-doping, the transition temperatures decrease. For x = 0.006, 0.013, 0.026, and 0.045, there are sharp anomalies at T$_N$ ≈ 127.6 K, 125.2 K, 117.6 K, and 106.2 K respectively. No features at lower temperatures, at T$^*$, were observed. For x = 0.049 and 0.054, there are peaks below T$_N$ ≈ 100 K and 97 K, respectively (Fig. 4, top inset). For these higher Mo-doping levels, the specific heat transition is broader, perhaps because 2 to 4 crystals were used to obtain a reasonable heat capacity signal. The plot of C/T versus T$^2$ dependence up to 10 K is shown in Fig. 4, bottom inset. The Sommerfeld-coefficient γ for all the compositions was estimated to range between ~ 6 to 13 mJ/(K$^2$mol). This weak change in γ with x is in contrast to strong increase in γ values for Ba(Fe$_{1-x}$Cr$_x$)$_2$As$_2$, with γ ≈ 30 mJ/(K$^2$ mol) for example found for x = 0.070.

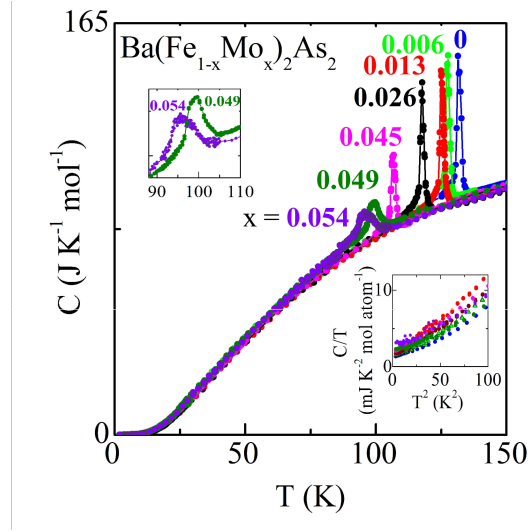

Fig. 4: For Ba(Fe$_{1-x}$Mo$_x$)$_2$As$_2$, temperature dependence of specific heat below 200 K. Bottom inset shows C/T versus T$^2$ form below 10 K.

Neutron diffraction experiments were performed on a Ba(Fe$_{1-x}$Mo$_x$)$_2$As$_2$ with x = 0.013 specifically to (a) confirm the coupled antiferromagnetic SDW order (T$_N$) to the orthorhombic phase (T$_o$) transition at ~ 125 K, (b) look for a possible ferromagnetism component below ~ 45 K (see Fig. 2b), and (c) look for the evidence of changes in magnetic and nuclear structure at ≈ 12.5 K (see Fig. 3a, bottom inset). The crystal (~0.01g) was measured at the High Flux Isotope Reactor at Oak Ridge National Laboratory, using the HB-1A and HB-3 triple-axis spectrometers. Due to the small crystal size, the HB-1A measurements were carried out with open collimations to gain more intensity, and with the (H K 0) plane aligned in the scattering plane. HB-3 measurements were performed with 48'-80'-80'-120' collimation with the (H 0 L) plane aligned in the scattering plane. The results are presented in Figure 5.

For x= 0.013, the tetragonal to orthorhombic structural transition was observed by measuring the temperature dependence of the (2 2 0)$_T$ nuclear peak (T refers to tetragonal basis), which shows extinction effects from the splitting into (4 0 0)$_O$ and (0 4 0)$_O$ reflections (O refers to orthorhombic basis) resulting in a significant change in the peak intensity at the transition. The order parameter of the magnetic transition from non-magnetic state to SDW order (with propagation vector (½ ½ 1)$_T$) was determined by the intensity of the strongest magnetic reflection: (½ ½ 3)$_T$; see Fig. 5a. In addition to the (2 2 0)$_T$ nuclear reflection (Fig. 5b), the temperature dependences of the nuclear (0 0 4)$_T$ and (1 0 1)$_T$, and the forbidden (1 0 0)$_T$ reflections were measured. Rocking curves at different temperatures for the (1 1 0)$_T$, (1 1 2)$_T$ and the aforementioned reflections were also measured to look for any structural effect, a ferromagnetic component or, as in the case of the similar hole-doped Cr-122, long range G-type antiferromagnetism with propagation vector (1 0 1)$_T$. Rocking curves of the (½ ½ 1)$_T$, (½ ½ 3)$_T$ and (½ ½ 7)$_T$ magnetic reflections were also measured at different temperatures in order to check for a change in the SDW magnetic structure.

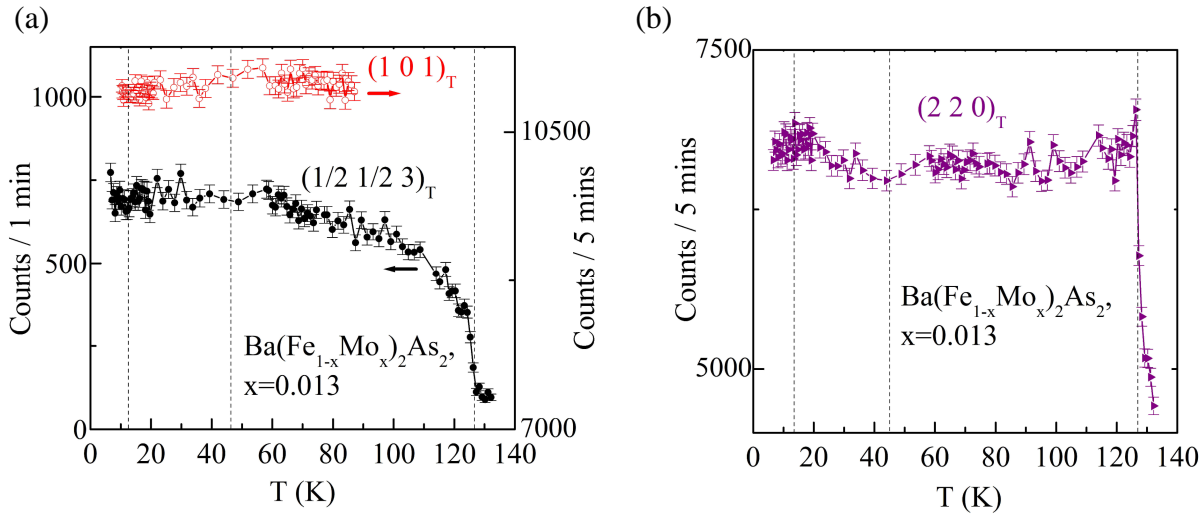

Fig. 5: (a) Temperature dependence of magnetic reflection (½,½,3)$_T$ (showing the onset of the magnetic transition at T$_N$ = 126 K) and nuclear reflection (1 0 1)$_T$ measured up to 87 K, showing the absence of a transition. (b) Temperature dependence of nuclear reflections (2 2 0)$_T$, showing extinction effects due to the structural transition at T$_0$ = 126 K.

Within the experimental sensitivity, no anomalies were observed in the temperature dependences (other than the expected transitions below ~ 126 K, see Fig. 5) nor in the rocking curves, seemingly indicating that no long-range structural, ferromagnetic or G-type antiferromagnetic transition happens at lower temperature. Changes in the magnetic structure (see Fig. 2b) may not have been observed due to the small size of the magnetic moments involved or because the ordering wave vector lies in a region of reciprocal space not probed by current measurements. Moreover, the coarse resolution and limited range of reciprocal space probed here may have prevented the observation of structural changes associated with the anomaly observed at $T^*$ (see Fig. 3a, bottom inset).

We looked for evidence of structural transition by performing low temperature powder x-ray diffraction (Cu K$\alpha_1$ radiation, Oxford PheniX cryostat) on powders ground from single crystals of x = 0.013 and x = 0.045. The angular range near the tetragonal (112) reflection [orthorhombic (202) and (022)] was carefully examined in these compositions. At 30 K, lattice parameters were determined from full-pattern LeBail refinements using FullProf program: For x = 0.045, the orthorhombic lattice constants are $a$ = 5.5955(6) Å, $b$ = 5.5604(8) Å, and $c$ = 12.984(1) Å; for x = 0.013, they are $a$ = 5.5733(3) Å, $b$ = 5.56100(2) Å, and $c$ = 12.9546(5) Å. For x= 0.013 and because of the feature at $T^* \approx$ 12.5 K (see Fig. 3a, bottom inset), powder x-ray diffraction was also collected at 11 K. We found that differences in the lattice constants between 30 and 11 K were found to be within experimental uncertainty.

**Conclusions**

Here we investigated the effects of 4$d$ molybdenum-doping in BaFe$_2$As$_2$. Based on the data above, an x-T phase diagram is proposed for the Ba(Fe$_{1-x}$Mo$_x$)$_2$As$_2$ system that is shown in Figure 6. According to temperature-dependent neutron diffraction and also anomalies seen in magnetic susceptibility, electrical resistivity, and heat capacity, there is the coupled SDW and structural transition for x = 0.013 at 125(1) K. The nature of magnetic transition temperature for all of x ≤ 0.066 is probably of the same antiferromagnetic SDW type. The suppression rate of magnetic transition temperature with x is the same as in 3$d$ Cr-doped 122, illustrating that T-x phase diagram for hole-doped 122 can be simply parameterized by x, similar to that found for electron-doped 122 systems. Further investigations are required to explain the anomalies in resistivity at $T^*$ for the lightly doped compositions of 0.006 ≤ x 0.028 in Ba(Fe$_{1-x}$Mo$_x$)$_2$As$_2$ and x = 0.022 in Ba(Fe$_{1-x}$Cr$_x$)$_2$As$_2$, and to understand the possible weak ferromagnetic component at low temperatures in Mo-122. The $T^*$ transition is x-dependent, similar way as $T_N$, and its electronic or magnetic origin may be related to the lack of bulk superconductivity in hole-doped Ba(Fe$_{1-x}$TM$_x$)$_2$As$_2$.

**Acknowledgements**

This work was partly supported by the Department of Energy, Basic Energy Sciences, Materials Sciences and Engineering Division and Scientific User Facilities Division. The Research at the High Flux Isotope Reactor of Oak Ridge National Laboratory was sponsored by the Scientific User Facilities Division, Office of Basic Energy Sciences, U.S. Department Of Energy.

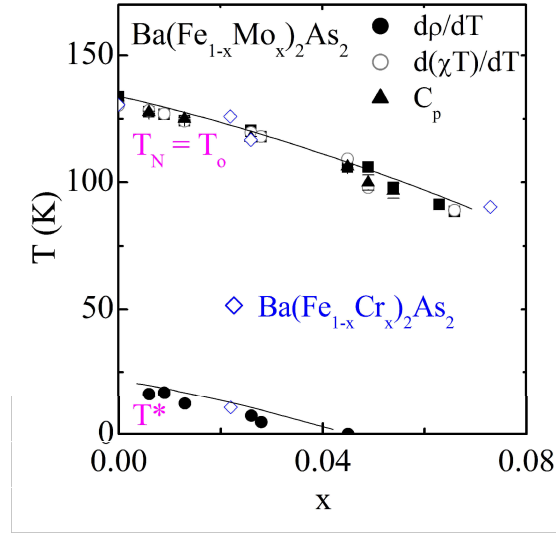

Figure 6: For Ba(Fe$_{1-x}$Mo$_x$)$_2$As$_2$, magnetic ordering temperature (T$_N$) versus Mo doping (x). Results from Ba(Fe$_{1-x}$Cr$_x$)$_2$As$_2$ (averaged results from [8] and here) are also added for comparison.